\documentstyle[12pt,epsfig]{article}
\def\cite#1{#1}

\def\thebibliography#1{\section*{References}\list
 {[\arabic{enumi}]}{\settowidth\labelwidth{[#1]}\leftmargin\labelwidth
 \advance\leftmargin\labelsep
 \usecounter{enumi}}
 \def\newblock{\hskip .11em plus .33em minus -.07em}
 \sloppy
 \sfcode`\.=1000\relax}

\title{New sum rules for nucleon and trinucleon total
photoproduction cross-sections\footnote{Contribution presented at
the PHOTON'03, April 7-11, 2003, Frascati (Roma), Italy}}

\author{S. Dubni\v cka$^1$, E. Barto\v s$^2$, E. Kuraev$^2$}
\date{\empty}

\begin{document}
\maketitle
\begin{center} {
$^{1}$  Inst. of Physics, Slovak Academy of Sciences, Bratislava,
Slovak Republic \\
$^{2}$ Bogoliubov Lab. of Theor. Physics, J.I.N.R., 141 980 Dubna,\\
        Moscow Region, Russia} \\
\end{center}

\vspace{0.2cm}

\begin{abstract}
Two new sum rules  are derived relating Dirac radii and anomalous
magnetic moments of the considered strongly interacting fermions
with the convergent integral over a difference of the total proton
and neutron, as well as $He^3$ and $H^3$, photoproduction
cross-sections.
 \vspace{1pc}
\end{abstract}

\maketitle

\section{INTRODUCTION}

A quarter of century ago Kurt Gottfried, by considering
high-energy electron-proton scattering and the nonrelativistic
quark model of hadrons, found \cite{Kurt67} a sum rule relating
the proton charge mean squared radius $\langle r^2_{Ep}\rangle$
and the proton magnetic moment $\kappa_p=1+\mu_p$ with the
integral over the total proton photoproduction cross-section
$\sigma_{tot}(\nu)$ in the following form
\begin{equation}
\int_0^{\infty}\frac{d\nu}{\nu}\sigma_{tot}(\nu)=\frac{\pi^2
\alpha}{m_p^2}[\frac{4}{3}m_p^2\langle r^2_{Ep}\rangle +
1-\kappa_p^2]\label{a1},
\end{equation}
where $\nu$ is the energy loss in the laboratory frame, $\alpha$
is the fine structure constant and $m_p$ is the proton mass.
Nowadays we know, that the Gottfried sum rule cannot be fulfilled
as the corresponding integral diverges due to the well known rise
of the total proton photoproduction cross-section at high
energies.

In this contribution by  a distinct way from the Gottfried
approach (utilizing analytic properties of the  forward Compton
scattering amplitude on the proton)  a proton sum rule similar to
(\ref{a1}) can be derived, however, suffering from the same
integral divergence alike as in (\ref{a1}) and suffering also from
the unknown left-hand cut contribution.

In order to avoid both of these shortcomings we are interested in
nucleon isodoublet (proton and neutron) simultaneously and derive
new sum rule, relating proton Dirac radius and anomalous magnetic
moments of the proton and the neutron to the integral over a
difference of the total proton and neutron photoproduction
cross-sections. Then, in the precision of the isotopic violation,
a mutual nullification of the left-hand cut proton and neutron
contributions,  as well as a convergence of the corresponding
integral in the sum rule, is achieved.

Because $He^3$ and $H^3$ belong to the same isodoublet, possess
the spin s=1/2 and as a result their electromagnetic  (EM)
structure is also completely described by Dirac and Pauli form
factors like proton and neutron, one can follow the same procedure
and derive a similar sum rule, relating $He^3$ and $H^3$ Dirac
radii and anomalous magnetic moments of $He^3$ and $H^3$ with the
convergent integral over a difference of the total $He^3$ and
$H^3$ photoproduction cross-sections.

\section{SUM RULE FOR DIFFERENCE OF PROTON AND NEUTRON TOTAL PHOTOPRODUCTION CROSS-SECTIONS}

As is well known, the proton and neutron are constituents of
atomic nuclei with the spin s=1/2, the EM structure of which is
completely described by two independent, e.g. Dirac $F_1(q^2)$ and
Pauli $F_2(q^2)$, form factors, defined by the relation
\begin{equation}
\langle N| J_{\mu}^{EM}|N \rangle = \bar
u(p')[\gamma_{\mu}F_1(q^2)+ 
 i \frac{\sigma_{\mu\nu}q^{\nu}}{2m_N}F_2(q^2)]u(p), \label{a2}
\end{equation}
where $J_{\mu}^{EM}=2/3\bar u\gamma_{\mu} u-1/3 \bar d
\gamma_{\mu} d-1/3 \bar s \gamma s$, $q^{\nu}=(p'-p)^{\nu}$ and
$\bar u(p')$, $u(p)$ are the free nucleon Dirac bi-spinors. Then
the asymptotic form of the very high energy elastic
electron-nucleon differential cross-section in the one-photon
approximation is
\begin{equation}
\frac{d\sigma^{e N\to e N}}{d{\bf \bf{q}}^2}= 4\frac{\pi
\alpha^2}{({\bf{q}}^2)^2}[F_1^2({\bf{q}}^2)+
\frac{{\bf{q}}^2}{4m_N^2}F_2^2(\bf{q}^2)]. \label{a3}
\end{equation}

Further, let us consider a very high energy peripheral
electroproduction process on nucleons
\begin{equation}
e^-(p_1) + N(p) \to e^-(p_1') + X, \label{a4}
\end{equation}
where the produced hadronic state $X$ is moving closely to the
direction of the initial nucleon. Its matrix element in the one
photon approximation takes the form
\begin{equation}
M = i\frac{\sqrt{4\pi \alpha}}{q^2} \bar u(p_1^{'})\gamma^\mu
u(p_1)<X \mid J^\nu \mid p>g_{\mu\nu}\label{a5}
\end{equation}
and $m^2_X = (p+q)^2$.

Now, by means of the method of equivalent photons \cite{Achie69},
examining the nucleon in the rest, the electron energy to be very
high and the small photon momentum transfer, one can express the
differential cross-section of the process (\ref{a4}) through
integral over the total nucleon photoproduction cross-section.

Really, applying to (\ref{a5}) the Sudakov expansion \cite{Sud56}
of the photon transferred 4-vector $q$
\begin{equation}\label{a6}
q=\beta_q \tilde{p}_1+\alpha_q\tilde{p}+q^{\bot}
\end{equation}
into the almost light-like vectors

\begin{equation}\label{a7}
\tilde{p}_1=p_1-p_1^2p/(2p_1p), \quad \tilde{p}=p-p^2p_1/(2p_1p),
\end{equation}
the Gribov representation \cite{Grib70} of the metric tensor

\begin{equation}
g_{\mu\nu}=g_{\mu\nu}^{\bot}+\frac{2}{s}(\tilde{p}_{\mu}\tilde{p}_{1\nu}+\tilde{p}_{\nu}\tilde{p}_{1\mu})
 \approx\frac{2}{s} \tilde{p}_\mu\tilde{p}_{1\nu}, \label{a8}
 \end{equation}
 where $s=(p_1+p)^2\approx 2p_1p\gg Q^2 = -q^2$
and transforming the phase space volume of the final state
suitably, one obtains
\begin{equation} \label{a9}
\frac{d\sigma^{e^-N\to
e^-X}}{d{\bf{q}}^2}=
\frac{\alpha{\bf{q}}^2}
{4\pi^2}\int\limits_{s_1^{th}}^\infty\frac{d
s_1}{s_1^2[{\bf{q}}^2+(m_es_1/s)^2]^2} Im
\tilde{A}(s_1,{\bf{q}})
\end{equation}
with $-t=Q^2={\bf{q}}^2$, $s_1=2qp=m^2_X+Q^2-m^2_N$ and
$Im\tilde{A}(s_1,\bf{q})$  to be  the imaginary part of the
forward Compton scattering amplitude on the nucleon
$\tilde{A}(s_1,\bf{q})$ with only the intermediate state $X$.

Finally, for the case of small photon momentum transfer squared in
(\ref{a9}) and by an application of the optical theorem
\begin{equation}\label{a10}
Im \tilde{A}(s_1,\bf{q})_{|_{\bf{q}^2\to
0}}=Im{A}(s_1,\bf{q})_{|_{\bf{q}^2\to 0}}=  4 s_1
\sigma_{tot}^{\gamma N\to X}(s_1)
\end{equation}
one obtaines the relation
\begin{equation} \label{a11}
{\bf{q}}^2\frac{d\sigma^{e^- N\to
e^-X}}{d{\bf{q}}^2}_{|_{{\bf{q}}^2\to 0}}=\frac{\alpha}
{\pi}\int\limits_{s_1^{th}}^\infty\frac{d s_1}{s_1}
\sigma_{tot}^{\gamma N\to X}(s_1)
\end{equation}
between the differential cross-section of the process (\ref{a4})
and the total nucleon photoproduction cross-section.

Next step is an investigation of analytic properties of the almost
forward Compton scattering amplitude $\tilde{A} (s_1,\bf{q})$ in
$s_1$-plane. They consist in one-nucleon intermediate state pole
at $s_1=-Q^2$, the right-hand cut starting at the pion-nucleon
threshold $s_1=-Q^2+2m_Nm_{\pi} +m_{\pi}^2$ and the $u_1$-channel
left-hand cut starting from $s_1= Q^2-8m_{N}^2$. Defining the path
integral $I$
\begin{equation}
I=\int_C\frac{d s_1}{({\bf
{q}}^2)^2}\frac{p_1^{\mu}p_1^{\nu}\tilde{A}_{\mu\nu}}{s_1^2}
\label{a12}
\end{equation}
\begin{figure}[htp] 
\centering
\includegraphics[scale=.35
]{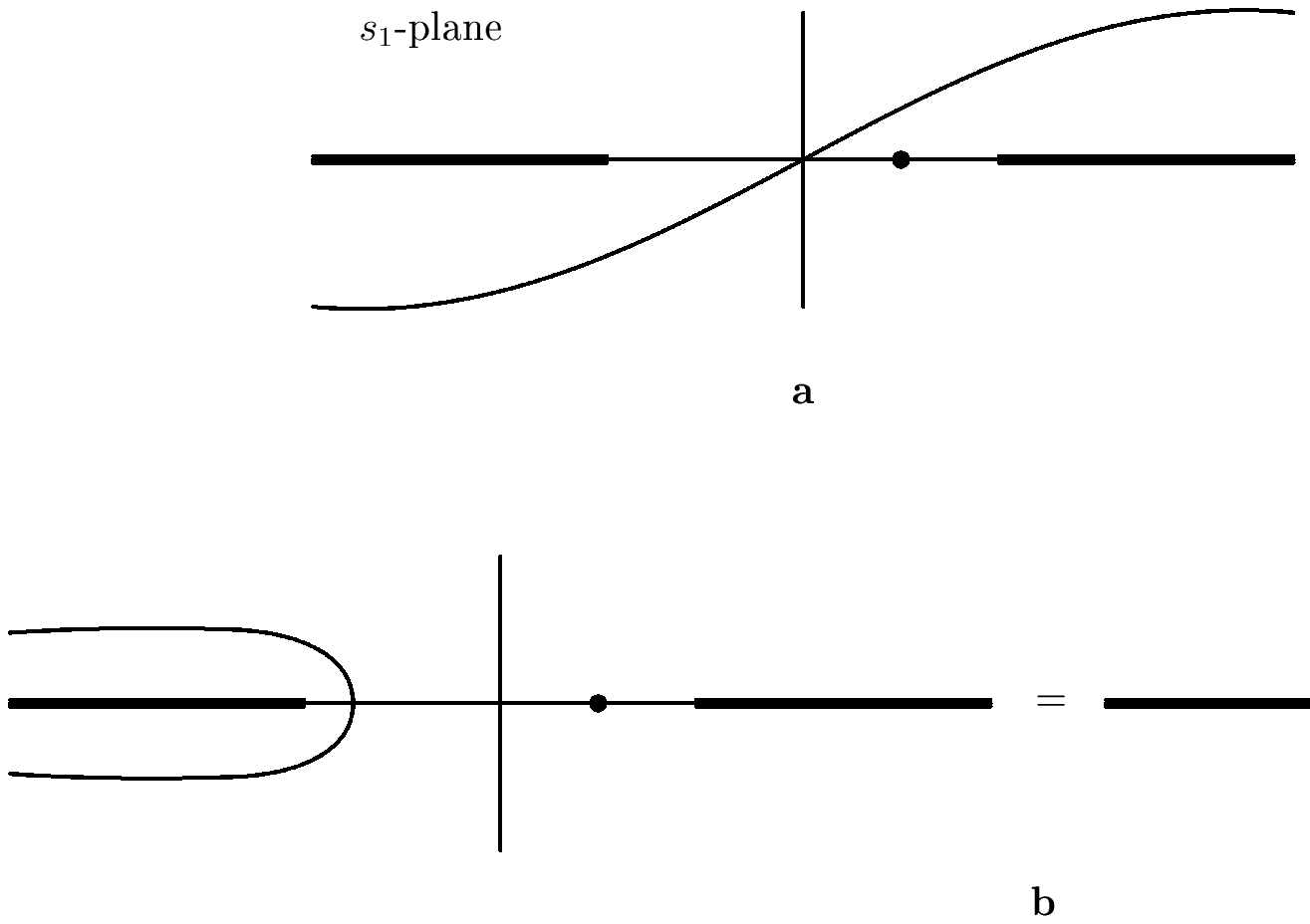} \caption{Sum rule interpretation in $s_1$
plane.}\label{fig1}
\end{figure}
from the gauge invariant light-cone projection
$p_1^{\mu}p_1^{\nu}\tilde{A}_{\mu\nu}$ of the part
$\tilde{A}_{\mu\nu}$ of the total Compton scattering tensor with
photon first absorbed and then emitted along the fermion line, in
$s_1$-plane, as presented in Fig.1 and once closing the contour
$C$ to upper half-plane, another one to lower half-plane, the
following sum rule
\begin{equation}\label{a13}
\pi \textmd{Res}=\int\limits_{r.h.}^\infty\frac{ds_1}
{s_1^2({\bf{q}}^2)^2} Im \tilde{A}(s_1,{\bf{q}})-
\int\limits_{l.h.}^{-\infty}\frac{ds_1} {s_1^2({\bf{q}}^2)^2}
Im \tilde{A}(s_1,{\bf{q}})
\end{equation}
appears with $\textmd{Res}$ to be the one-nucleon intermediate
state pole contribution. Then, taking into account (\ref{a3}) and
(\ref{a9}) and considering proton and neutron separately, one
comes from (\ref{a13}) (for more detail see Appendix D in
\cite{Kura81}) to the sum rules for the proton
\begin{equation}\label{a14}
1-F_{1p}^2(-Q^2)-\frac{Q^2}{4m_p^2}F_{2p}^2(-Q^2)=
\frac{(Q^2)^2}{\pi\alpha^2} \frac{d\sigma^{e^-p\to
e^-X}}{dQ^2}+ LCC_p
\end{equation}
and for the neutron
\begin{equation}\label{a15}
-F_{1n}^2(-Q^2)-\frac{Q^2}{4m_n^2}F_{2n}^2(-Q^2)=
=\frac{(Q^2)^2}{\pi\alpha^2} \frac{d\sigma^{e^-n\to
e^-X}}{dQ^2}+ LCC_n,
\end{equation}
where the obreviations $LCC_p$ and $LCC_n$ denote the left-hand
cut contributions in (\ref{a13}) for proton and neutron,
respectively.

A similar proton sum rule to (\ref{a1}) can be now found taking a
derivative of both sides in (\ref{a14}) according to $Q^2$ at
$Q^2\to 0$ and substituting the relation (\ref{a11}) for proton.

However, in order to avoid the abovementioned problems we
substract (\ref{a15}) from (\ref{a14}) and as a result (due to a
small isotopic invariance violation in the EM interactions) one
achieves a mutual annulation of the left-hand cut proton and
neutron contributions as follows
\begin{eqnarray}\label{a16}
&&1-F_{1p}^2(-Q^2)+F_{1n}^2(-Q^2)-
\frac{Q^2}{4m_p^2}F_{2p}^2(-Q^2)+
\frac{Q^2}{4m_n^2}F_{2n}^2(-Q^2)= \nonumber \\
&=&\frac{(Q^2)^2}{\pi\alpha^2} \Bigg(\frac{d\sigma^{e^-p\to
e^-X}}{dQ^2}- \frac{d\sigma^{e^-n\to e^-X}}{dQ^2}\Bigg).
\end{eqnarray}
Moreover, taking a derivative  according to $Q^2$ of both sides in
(\ref{a16}) and  employing the relation (\ref{a9}),  one comes for
$Q^2\to 0$ to the  new sum rule relating Dirac mean squared radius
and anomalous magnetic moments of the proton and the neutron to
the integral over  a difference of the total proton and neutron
photoproduction cross-sections
\begin{equation}\label{a17}
\frac{1}{3}\langle r_{1p}^2 \rangle
-\frac{\mu_p^2}{4m_p^2}+\frac{\mu_n^2}{4m_n^2}=
\frac{1}{\pi^2\alpha}\int\limits_{\omega_N}^\infty
\frac{d\omega} {\omega}\big[\sigma_{tot}^{\gamma p\to X}(\omega)-
\sigma_{tot}^{\gamma n\to X}(\omega)\big] 
\end{equation}
with $\omega_N=m_{\pi}+m_{\pi}^2/2m_N$, in which just a mutual
cancellation of the rise of the latter cross sections for $\omega
\to \infty$ is achieved.

\section{SUM RULE FOR He$^3$ and  H$^3$ TOTAL PHOTOPRODUCTION
CROSS-SECTIONS}

Isodoublet of nuclei $He^3$ and $H^3$ has, like nucleons, spin
s=1/2 and the elastic and inelastic scatterings of electrons on
these nuclei are described by similar formulas like nucleons. Only
normalizations of the corresponding Dirac and Pauli form factors
are slightly different
\begin{eqnarray}
F_{1He^3}(0)=2; \quad &&F_{2He^3}(0)=\mu_{He^3}; \label{a18} \\
F_{1H^3}(0)=1; \quad && F_{2H^3}(0)=\mu_{H^3},\nonumber
\end{eqnarray}
where $\mu_{He^3}$ and $\mu_{H^3}$ are anomalous magnetic moments
of $He^3$ and $H^3$, respectively. So, repeating the same
procedure with the latter nuclei one obtaines a new sum rule of
the following form
\begin{eqnarray}
&&\frac{1}{3}\big[4\langle r_{1He^3}^2 \rangle
-\langle r_{1H^3}^2 \rangle\big]-
\frac{\mu_{He^3}^2}{4m_{He^3}^2}+\frac{\mu_{H^3}^2}{4m_{H^3}^2}= \label{a19} \\
&=&\frac{1}{\pi^2\alpha}\int\limits_{\omega_N}^\infty
\frac{d\omega} {\omega}\big[\sigma_{tot}^{\gamma He^3\to
X}(\omega)- \sigma_{tot}^{\gamma H^3\to X}(\omega)\big] \nonumber
\end{eqnarray}
 relating Dirac radii and anomalous
magnetic moments of $He^3$ and $H^3$ with the convergent integral
over a difference of the total $He^3$ and $H^3$ photoproduction
cross-sections.

\section{CONCLUSIONS}

Considering the very high energy elastic and inelastic
electron-nucleon and electron trinucleon scattering with a
production of a hadronic state $X$ moving closely to the direction
of initial hadrons. Then utilizing analytic properties of the 
forward Compton scattering amplitudes on considered strongly
interacting particles, for the case of small transfered momenta
new sum rules were derived, relating Dirac radii and anomalous
magnetic moments of these fermions with the convergent integral
over a difference of the total proton and neutron, as well as
$He^3$ and $H^3$, photoproduction cross-sections.

The work was partly supported by Slovak Grant Agency for Sciences,
Grant 2/1111/22  (S.D.).

\end{document}